\documentclass[journal]{IEEEtran}
\bstctlcite{IEEEexample:BSTcontrol}
\usepackage{ragged2e}
\usepackage{array}
\usepackage{amssymb}
\usepackage[noadjust]{cite}

\ifCLASSINFOpdf
 
\else
  
\fi

\usepackage{scalerel}
\usepackage{tikz}
\usetikzlibrary{svg.path}

\definecolor{orcidlogocol}{HTML}{A6CE39}
\tikzset{
  orcidlogo/.pic={
    \fill[orcidlogocol] svg{M256,128c0,70.7-57.3,128-128,128C57.3,256,0,198.7,0,128C0,57.3,57.3,0,128,0C198.7,0,256,57.3,256,128z};
    \fill[white] svg{M86.3,186.2H70.9V79.1h15.4v48.4V186.2z}
                 svg{M108.9,79.1h41.6c39.6,0,57,28.3,57,53.6c0,27.5-21.5,53.6-56.8,53.6h-41.8V79.1z M124.3,172.4h24.5c34.9,0,42.9-26.5,42.9-39.7c0-21.5-13.7-39.7-43.7-39.7h-23.7V172.4z}
                 svg{M88.7,56.8c0,5.5-4.5,10.1-10.1,10.1c-5.6,0-10.1-4.6-10.1-10.1c0-5.6,4.5-10.1,10.1-10.1C84.2,46.7,88.7,51.3,88.7,56.8z};
  }
}

\newcommand\orcidicon[1]{\href{https://orcid.org/#1}{\mbox{\scalerel*{
\begin{tikzpicture}[yscale=-1,transform shape]
\pic{orcidlogo};
\end{tikzpicture}
}{|}}}}
\usepackage{hyperref} 

\begin{document}

\title{56 GBaud PAM-4 100 km Transmission System with Photonic Processing Schemes}

\author{Irene Est\'ebanez \orcidicon{0000-0002-0130-8774},
        Shi Li \orcidicon{0000-0001-7237-8932}, Janek Schwind, Ingo Fischer \orcidicon{0000-0003-1881-9842}, Stephan Pachnicke \orcidicon{0000-0001-7321-7938}, \textit{Senior Member, IEEE} and Apostolos Argyris \orcidicon{0000-0003-2847-7719} 

\thanks{This work was supported by MINECO (Spain), through project TEC2016-80063-C3 (AEI/FEDER, UE) and by the Spanish State Research Agency, through the Severo Ochoa and Mar\'ia de Maeztu Program for Centers and Units of Excellence in R$\&$D (MDM-2017-0711). The work of I. Est\'ebanez  has  been  supported  by  MICINN,  AEI,  FEDER and the  University  of  the  Balearic  Islands  through  a predoctoral  fellowship  (MDM-2017-0711-18-2). The work of A. Argyris was supported by the Conselleria d’Innovació, Recerca i Turisme del Govern de les Illes Balears and the European Social Fund.}
\thanks{I. Est\'ebanez, I. Fischer, and A. Argyris are with Instituto de Física
Interdisciplinar y Sistemas Complejos IFISC (CSIC-UIB), Campus UIB,
Palma de Mallorca, 07122, Spain (e-mail: irene@ifisc.uib-csic.es;
ingo@ifisc.uib-csic.es; apostolos@ifisc.uib-csic.es).}
\thanks{J. Schwind was jointly with Instituto de Física Interdisciplinar y Sistemas
Complejos IFISC (CSIC-UIB), Campus UIB, Palma de Mallorca, 07122,
Spain, and Institute of Applied Physics, University of Münster, Münster,
48149, Germany (e-mail: janekschwind@gmx.de).}
\thanks{S. Li and S. Pachnicke are with the Chair of Communications, Kiel University, Kiel, 24143, Germany (e-mail: shi.li@tf.uni-kiel.de; stephan.pachnicke@tf.uni-kiel.de).}}

\noindent\fbox{%
    \parbox{0.95\textwidth}{%
        \large{© 2021 IEEE.  This work has been accepted for publication in the Journal of Lightwave Technology, with DOI: 10.1109/JLT.2021.3117921. Personal use of this material is permitted.  Permission from IEEE must be obtained for all other uses, in any current or future media, including reprinting/republishing this material for advertising or promotional purposes, creating new collective works, for resale or redistribution to servers or lists, or reuse of any copyrighted component of this work in other works.
        }
    }%
}
\maketitle
\begin{abstract}
Analog photonic computing has been proposed and tested in recent years as an alternative approach for data recovery in fiber transmission systems. Photonic reservoir computing, performing nonlinear transformations of the transmitted signals and exhibiting internal fading memory, has been found advantageous for this kind of processing. In this work, we show that the effectiveness of the internal fading memory depends significantly on the properties of the signal to be processed. Specifically, we demonstrate two experimental photonic post-processing schemes for a 56 GBaud PAM-4 experimental transmission system, with 100 km uncompensated standard single-mode fiber and direct detection. We show that, for transmission systems with significant chromatic dispersion, the contribution of a photonic reservoir’s fading memory to the computational performance is limited. In a comparison between the data recovery performances between a reservoir computing and an extreme learning machine fiber-based configuration, we find that both offer equivalent data recovery. The extreme learning machine approach eliminates the necessity of external recurrent connectivity, which simplifies the system and increases the computation speed. Error-free data recovery is experimentally demonstrated for an optical signal to noise ratio above 30 dB, outperforming an implementation of a Kramers-Kronig receiver in the digital domain. 
\end{abstract}


\IEEEpeerreviewmaketitle

\section{Introduction}

\IEEEPARstart{C}{onceptual} simplifications in both feed-forward and recurrent neural network approaches have been gaining appeal in recent years. Two of the most approved random-projection techniques in these two approaches are the extreme learning machines (ELMs) \cite{huang2006} and the reservoir computing (RC) \cite{jaeger2001echo,maass2002real,jaeger2004harnessing}. An unquestionable reasoning is that they ease diverse hardware implementations in various physical substrates \cite{Ortin2015,van2017,tanaka2019,chembo2020,genty2020}. While gradient descent methods are needed to train deep feed-forward neural networks, the former use unaltered structures to generate their nonlinear responses. These responses are then used to train linear classifiers at the readout layer, targeting prediction, classification, or time-dependent signal processing. In the recent years, hardware configurations that are based on ELM and/or RC have been numerically investigated and/or experimentally tested, aiming at data recovery in fiber transmission systems \cite{Argyris2017,Argyris2018,Argyris2019,Shi2019,Estebanez2020,Sorokina2018,Sorokina2019dual,Katumba2019,Sorokina2020dispersion} and header recognition tasks \cite{vandoorne2014,qin2017,katumba2018,zhao2018,freiberger2019}. RC topologies have been also considered as building blocks at the digital signal processing stage of data decoders \cite{DaRos2020,Ranzini2021}. Many of the used topologies follow the time delay RC (TDRC) simplification, where a feedback loop is used to introduce the recurrency in the reservoir \cite{Appeltant2011}. In fiber transmission systems and tasks related to data recovery, the reservoir recurrency offers an internal memory that may prove beneficial for the computation performance \cite{Argyris2018}. However, as we show in this study, this is not always the case.

Here we present an experimental data recovery task in a fiber transmission system, where an ELM can perform similarly well as a TDRC configuration. We built experimentally a 5-channel, 56 GBaud, dense wavelength division multiplexing (DWDM), 100 km transmission system with direct-detection (IM/DD), and 4-level pulse amplitude modulation (PAM-4) with single side-band (SSB) encoding. This transmission system has been evaluated in the recent past \cite{Shi2019} by considering only a numerical implementation of a photonic TDRC, with promising results. However, the impact of the reservoir and the inherent fading memory on the final performance was not addressed in that work. Transmission systems like this, which suffer from extended chromatic dispersion, may benefit only slightly from the recurrent connectivity of the TDRC. Thus, here, we present an experimental implementation of a photonic TDRC / ELM configuration, based on the design presented in \cite{Ortin2015}, to process the data obtained from this transmission system \cite{Shi2019}. An ELM configuration simplifies the processing system significantly, by eliminating the need for an optical feedback loop. In this way, the overall computational speed is defined at the input layer pre-processing stage and not by the time delay of the photonic reservoir \cite{Argyris2018}. Finally, we compare the TDRC and the ELM decoding performance with the one obtained from Kramers-Kronig (KK) reception as implemented in \cite{Ohlendorf2018,Shi2019} by digital signal processing (DSP). 

\section{Experimental transmission system}

\begin{figure}[h]
   \centering
        \includegraphics[width=0.65\linewidth]{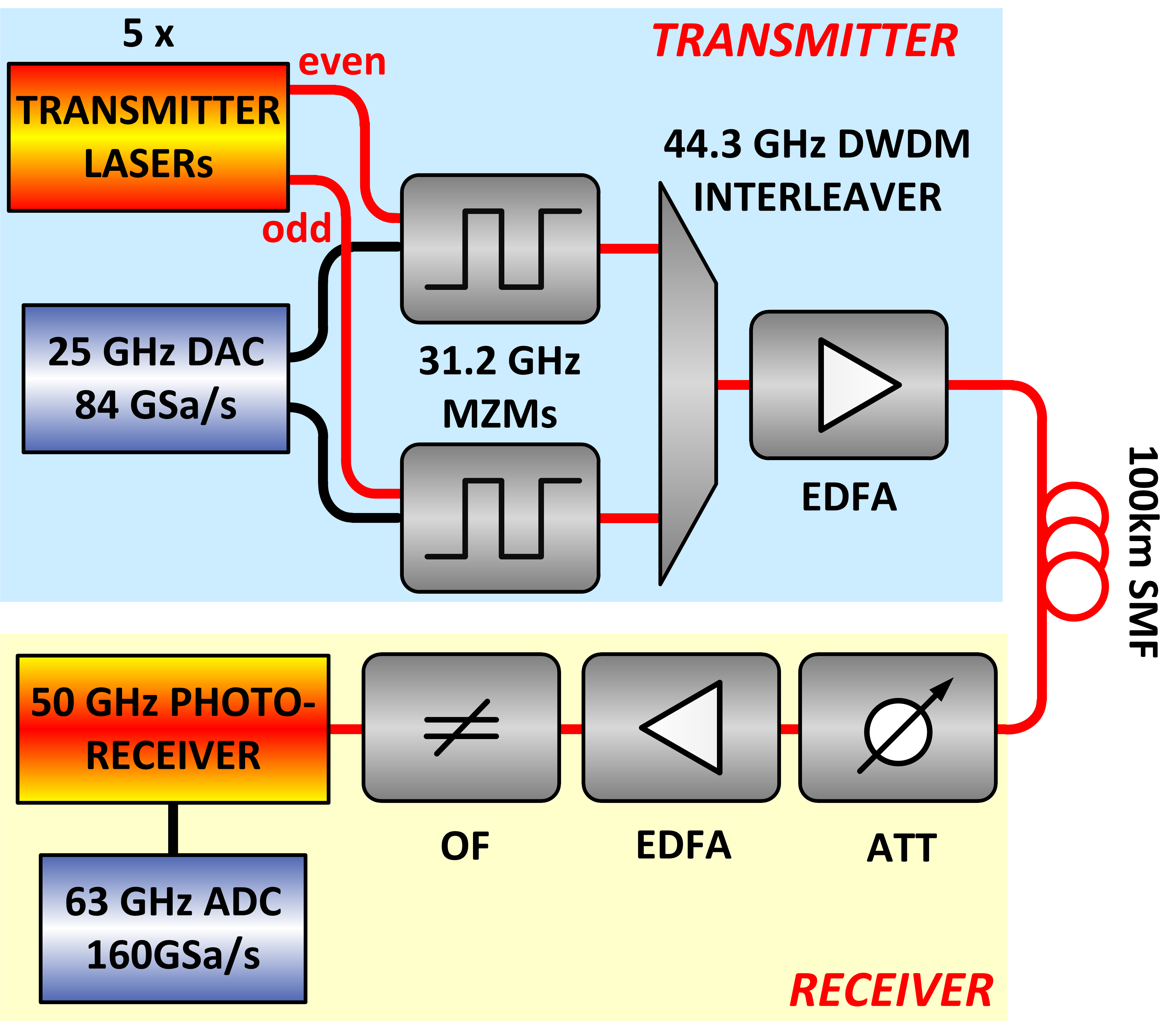}
    \caption{Experimental DWDM incoherent transmission system with PAM-4 encoding at 56GBaud and $100$ km standard single mode fiber transmission.}
    \label{fig:figure1}
\end{figure}

Fig. \ref{fig:figure1} shows the 100 km DWDM transmission implementation, as demonstrated in \cite{Shi2019,Ohlendorf2018}. The random Gray-coded PAM-4 data is pulse-shaped into root-raised cosine with a roll-off factor of $\beta = 0.1$. To compensate for the low-pass characteristics of the digital-to-analog converter (DAC) and the Mach-Zehnder modulators (MZMs), the signal is pre-emphasized and additionally optimized for minimum signal-signal beat interference while upholding the minimal phase condition by adjusting the carrier to signal power ratio (CSPR). Five independent channels are optically generated using two linearly driven ($0.25\cdot V\pi$) MZMs and five external cavity lasers (ECLs). For the single sideband (SSB) transmission, the ECLs are frequency detuned by 24.5 GHz from the channel center and subsequently filtered by the DWDM multiplexer (interleaver). After amplification, the optical signal is transmitted over 100 km uncompensated standard single-mode fiber (SSMF). The OSNR is varied by noise loading after transmission from a second amplification unit. The highest OSNR obtained by this system was 36.6 dB and was achieved for 6.5 dBm launched optical power per optical DWDM channel. For such optical power levels, the transmitted signal is significantly distorted by nonlinear phenomena, such as the self-phase and the cross-phase modulation. After optical filtering of the center channel (channel under test) and direct detection (DD), the signal is synchronized and re-sampled from the analog-to-digital (ADC) sample rate of 160 GSa/s to 2 samples per symbol (SpS). The effective number of bits for the DAC and ADC is 5.5. This signal is used to feed the photonic processing schemes of TDRC and ELM, as we describe in the next session. However, we also process this signal, by using a conventional DSP approach; a KK receiver scheme, in which the phase of the signal is recovered using the KK relation \cite{Shi2019}. Further noise reduction and compensation of chromatic dispersion (CD) are performed by a matched filter and a feed-forward equalizer (FFE) using 48 taps. After demapping, the BER is calculated by error counting and its performance is compared to the BER obtained by the photonic processing schemes.

\begin{figure}[t]
   \centering
        \includegraphics[width=1\linewidth]{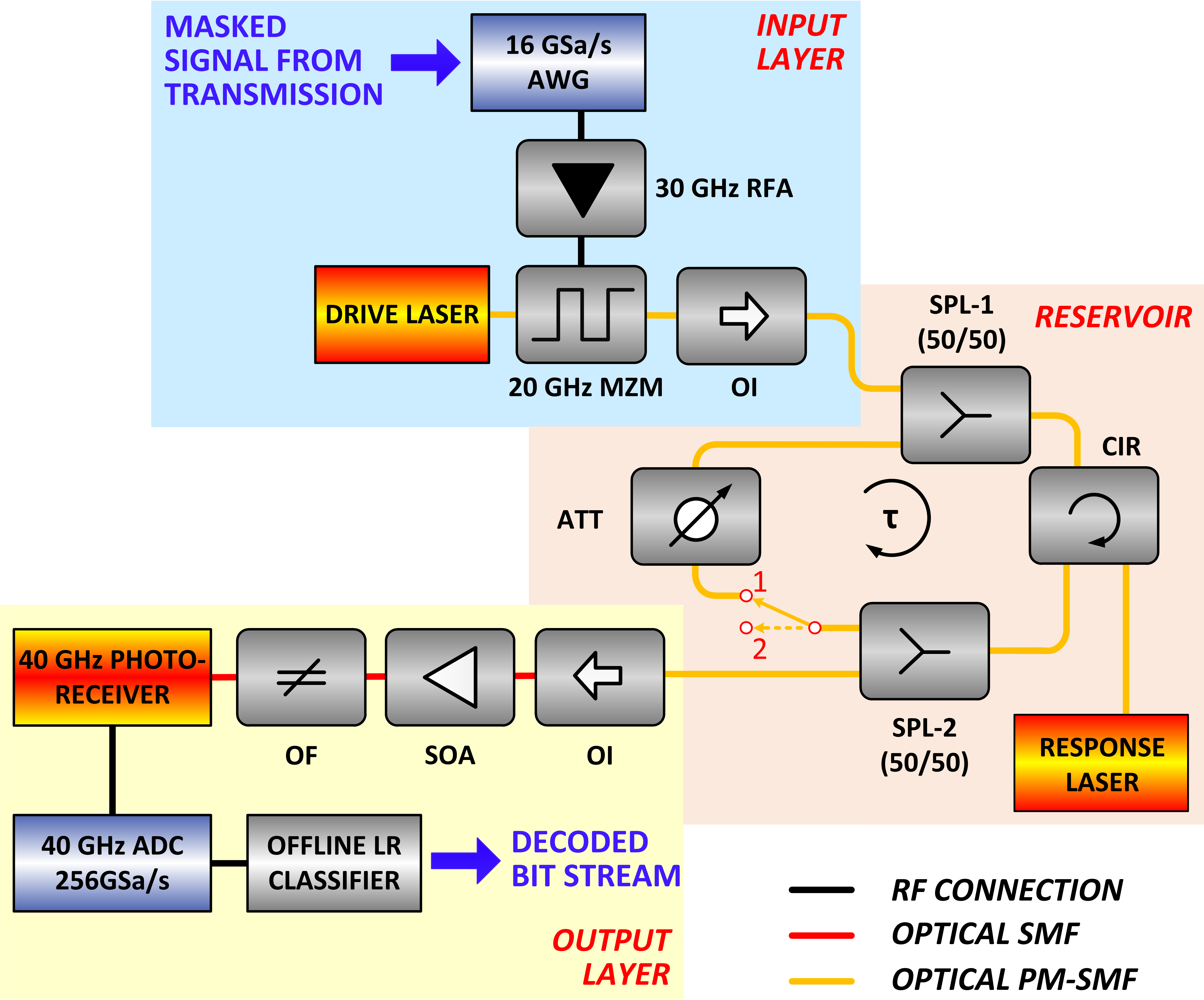}
    \caption{Experimental photonic TDRC and ELM setup. When the switch is in position 1, the optical feedback loop is connected and the TDRC scheme is activated. When the switch is in position 2 (floating connection), the feedback loop is open and an ELM scheme is activated.}
    \label{fig:figure2}
\end{figure}

\section{Experimental photonic reservoir and extreme learning machine}

\subsection{Photonic Reservoir}

\begin{figure}[t]
   \centering
        \includegraphics[width=0.9\linewidth]{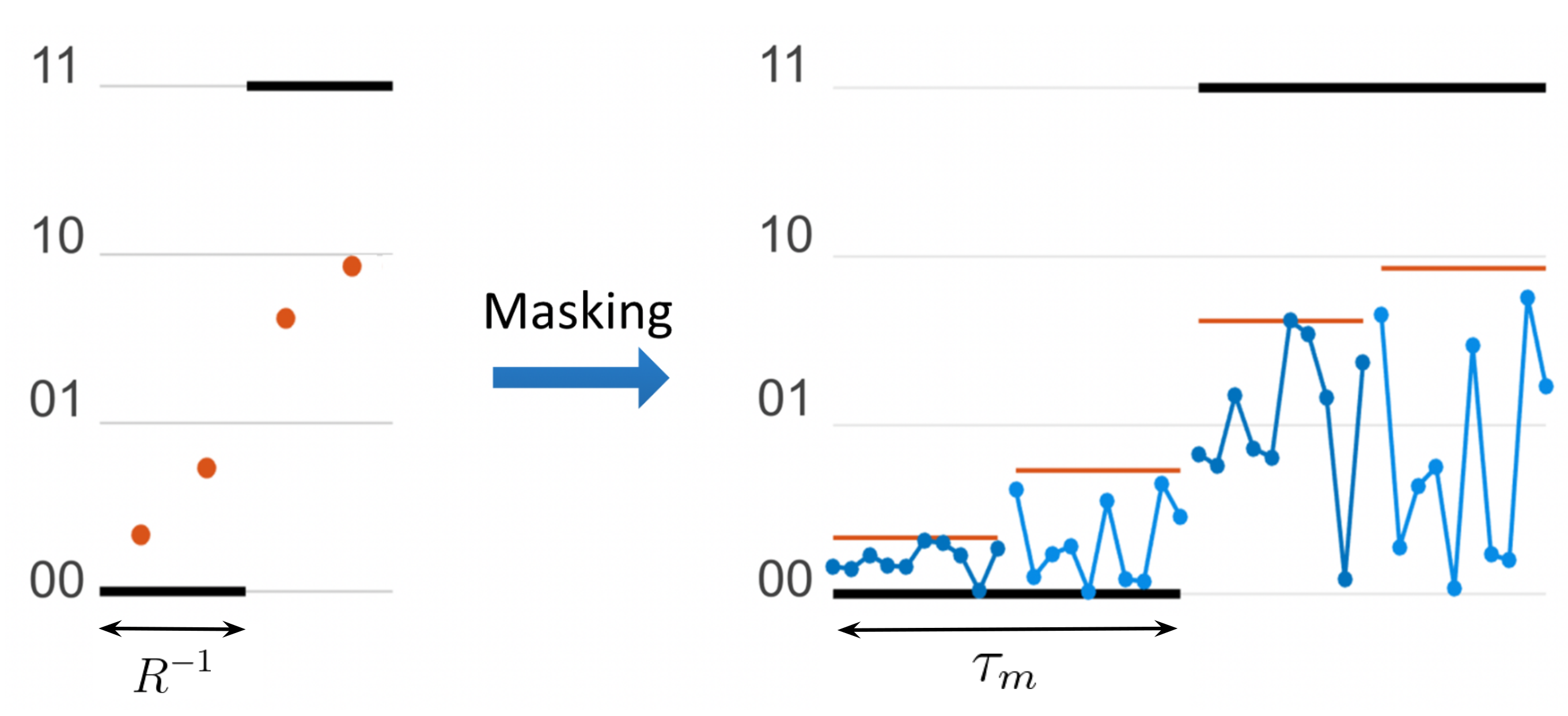}
    \caption{Masking of the PAM$-4$ transmission signal before photonic processing. The two samples per symbol (red dots) obtained from the transmission detection system expand to $N$ samples after masking (blue dots). The masking process introduces a computational speed penalty of $\tau_{m}/R$, where $\tau_{m} = N \cdot \theta$ is the mask duration and $R$ is the encoding data rate.}
    \label{fig:figure3}
\end{figure}

The recorded signal from the transmission system's ADC, is used as the signal of the input layer of the single-mode fiber-based (SMF) experimental TDRC scheme, shown in Fig. \ref{fig:figure2}. The signal from the transmission is multiplied offline with a random sequence (mask) to expand its dimensionality \cite{Appeltant2011}, as shown in Fig. \ref{fig:figure3}. Each symbol is multiplied with an $N$-valued mask, where $N$ is the mask dimension. The mask pattern is a random sequence drawn from a uniform distribution $[0,1]$ and is preserved for all processed symbols. Since each symbol consists of 2 samples, we always consider $N$ an even number, so that every sample of each symbol is masked with $N/2$ values. As shown in Fig. \ref{fig:figure2}, the masked signal is uploaded into an arbitrary waveform generator (AWG) and transformed into an electrical modulation signal with a sampling rate of 16 GSa/s. This defines a sampling duration for the mask values $-$ and an equal virtual node time separation $-$ of $\theta = 62.5$ ps. The AWG's output is amplified by a $30$ GHz broadband RF amplifier (RFA) and modulates the optical carrier of the drive laser (SL), via a $20$ GHz MZM. This optical signal carries the information to be injected into the response laser. The injection power is tuned via the bias current of the injection laser. The coupling of light into the response laser is realized via a 50/50 optical splitter (SPL$-1$) and a $3$-port optical circulator (CIR). The average optical injection power entering into the response laser is measured $96 \mu W$. 

When connecting the switch to position 1 in  Fig. \ref{fig:figure2}, we define a photonic reservoir with a fiber delay loop ($\tau=24.5$ ns), which is shorter compared to the ones previously presented \cite{Argyris2019,Shi2019,Estebanez2020,Sorokina2018,Sorokina2019dual}. The reservoir structure is implemented by using polarization-maintaining (PM) SMF components to establish a robust operation of the optical injection system over time. The response laser is biased below the solitary threshold at $10.1$ mA ($I_{th}=10.2$ mA), while the wavelength of its optical emission is $1545.5$ nm. The drive DFB SL is chosen to emit at a close wavelength and is tuned by controlling its operating temperature. The frequency detuning between the drive and the response laser $\Delta f = f_{d}-f_{r}$ is changed with a resolution of $0.01$ K, which corresponds to $\sim 125$ MHz and is measured with a $10$ MHz resolution optical spectrum analyzer. The frequency difference $\Delta f$ is critical since it determines the nonlinear transformation that is performed by the response laser to the input sequence. In the TDRC scheme, the optical feedback loop includes an optical attenuator (ATT) that controls the optical feedback strength and a 50/50 splitter (SPL$-2$) that leads to the switch and the possibility to close the feedback loop.

When connecting the switch to position 2 in Fig. \ref{fig:figure2}, we eliminate the recurrency of the TDRC scheme and thus the fading memory at the characteristic time $\tau$. Thus, we establish an ELM configuration \cite{Ortin2015}, in which the masked input signal undergoes a nonlinear transformation when passing through the response laser and consecutively is directed to the output port of the SPL$-2$. 

Eventually, at the output port of SPL$-2$, the optical signal is amplified by a semiconductor optical amplifier (SOA) and filtered by a tunable optical filter (OF). An optical isolator (OI) minimizes the back-reflections ($\sim 60$ dB) to the response laser. The output optical signal is detected by a $40$ GHz photoreceiver and received by a $256$ GSa/s ADC of a $40$ GHz real-time oscilloscope. Multiple averages of the same processed input time stream are used to improve the signal-to-noise ratio of the detected signal. Finally, the recorded time-series are used to train a linear (LR) classifier and to evaluate the classification performance of the received data versus the initial encoding data stream.

\begin{figure}[h]
   \centering
        \includegraphics[width=1\linewidth]{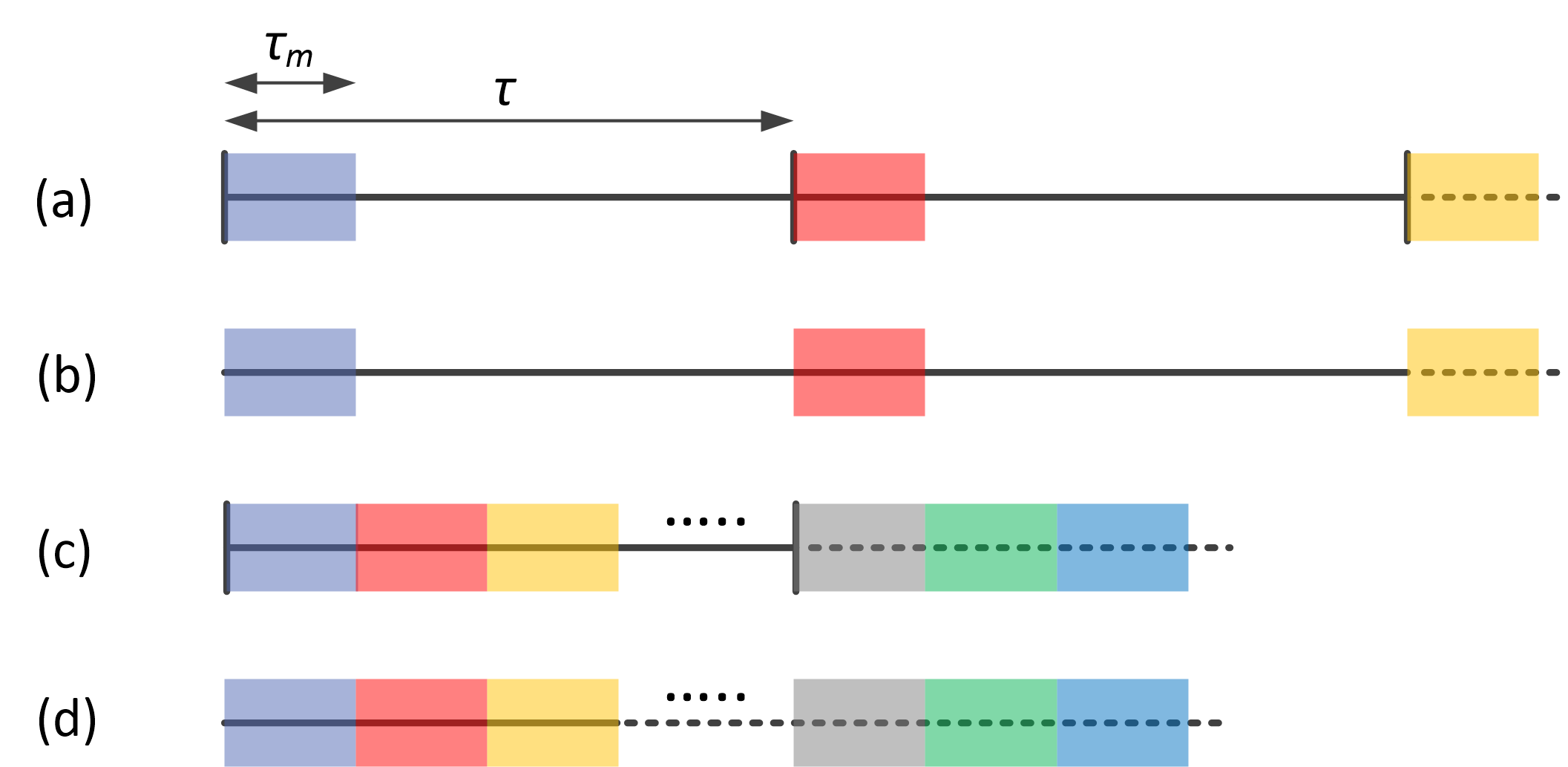}
    \caption{Schematic of the different ways to introduce the masked signal into the reservoir system. Black vertical lines represent the presence of the delay loop ($\tau$) in the reservoir cavity. Different color data packets correspond to the consecutive symbol time duration after masking. (a) One masked symbol processed within one reservoir time delay. (b) Same as (a) but in absence of the feedback loop. (c) $n$ sequential masked symbols that fit within one reservoir time delay. (d) Same as (c) but in absence of the feedback loop.}
    \label{fig:figure4}
\end{figure}

\subsection{Input encoding methods}
Previous experimental implementations of a fiber-based TDRC approach have been successful in data recovery tasks from fiber transmission systems \cite{Argyris2017,Argyris2018,Argyris2019}. But the processing of one symbol of encoded information was always assigned to one reservoir time delay $\tau$, defining, thus, a processing speed of $1/\tau$. The way that the input information is processed defines not only the final computational performance but also the speed at which one unit of information is processed. By considering different approaches of input data encoding and the presence or not of the reservoir's feedback loop, we aim at the most efficient way to process the information of this transmission system, in terms of speed and performance. In Fig. \ref{fig:figure4} we show the different approaches to encode each masked symbol with a duration of $\tau_{m}$, in a time delay of duration $\tau$.
\begin{enumerate}
\item[(a)]One masked symbol encoded per time delay $\tau$ in TDRC configuration. The inertia from the transient effects introduces connectivity between the neighboring nodes within a masked symbol duration. However, inertia connectivity between symbols is absent. Subsequent neighboring masked symbols are connected through optical feedback. The processing speed is in this case $1/\tau$. 
\item[(b)]One masked symbol encoded per time delay $\tau$ in ELM configuration. The neighboring symbols are not connected either by transient inertia of the previous state or by optical feedback. Only the inertia from the transient effects introduces connectivity between the neighboring nodes. Although the time delay $\tau$ has no physical definition in the ELM configuration, we investigate this case only to compare with case (a) and identify the contribution of the optical feedback connectivity. The processing speed is again $1/\tau$. 
\item[(c)]Multiple masked symbols encoded per time delay $\tau$ in TDRC. Here the inertia from the transient effects connects the neighboring virtual nodes but also the neighboring masked symbols. In the presence of feedback in the TDRC operation, there is additional connectivity after every $n$ masked symbols, where $n$ is the number of symbols that fit within a time delay $\tau$. In this case, the processing speed is $ n/\tau$. 
\item[(d)]Multiple masked symbols encoded in ELM. The connectivity between neighboring nodes and masked symbols originates only from inertia. In this case, the processing speed is $1/\tau_{m}$. 
\end{enumerate}
These connectivity properties are summarized in Table I. In all cases, however, the training of the linear classifier at the output layer can use as many responses as needed from neighboring symbols (taps) to optimize its performance. This external memory introduces only latency, and no speed penalty, to the computation.

\begin{table}[t]
\caption{Types of connectivity for the different input encoding methodologies.}
\label{table_1}
\centering
\begin{tabular}{ m{1.5cm}  m{1.72cm} m{1.72cm} m{1.72cm}}
\hline
\hline
Encoding methodology & Connectivity through inertia between transient states & Connectivity through feedback loop   & Connectivity through inertia between symbols \\
\hline
(a) & $\checkmark$& $\checkmark^{1}$ & $\times$\\
(b) & $\checkmark$ & $\times$ & $\times$\\
(c) & $\checkmark$ & $\checkmark^{2}$ & $\checkmark$\\
(d) & $\checkmark$ & $\times$ & $\checkmark$\\
\hline
\hline
\end{tabular}

\justify
1: the reservoir’s feedback loop introduces connectivity between subsequent masked symbols. 2: the reservoir's feedback loop introduces connectivity between masked symbols that have a distance of $n$.
\end{table}

In the fiber-based implementation of the TDRC system (Fig. \ref{fig:figure2}), the selected $\theta$ value of $62.5$ ps keeps the response laser in a transient state, while we can define in total $N_{t}=\tau / \theta = 392$ virtual nodes along with the delay. But this dimensionality increase reduces the computational speed. The $24.5$ ns fiber delay is still much longer than the one requested for the mask dimensionality expansion $N$ of our task. Initially, we consider $N = 20$ virtual nodes, thus $\tau_{m}= N \cdot \theta = 1.25$ ns. In the encoding case of Fig. \ref{fig:figure4}c, the number of masked symbols $n$ that fit in a time delay $\tau$ is the quotient of the Euclidean division $\tau / \tau_{m}$. 

\subsection{Classifier properties}
The virtual nodes' responses acquired at the output of the reservoir are used to train a ridge regression classifier, with a ridge parameter equal to 0.01. For the specific computing task, we identified an optimal training size of $16500$ symbols, while $12000$ symbols were used as the testing set. The training and testing data sets were separated with a buffer data set of $500$ symbols, which were not used, to eliminate any bias in the computation. For the minimization of the error rate, we consider up to $61$ taps (nonlinear responses of our photonic system that correspond to neighboring symbols), and select that tap size that provides us with the lowest bit error rate (BER). In all the investigated operating conditions and system configurations, we find that the optimal number of taps lies between $31$ and $51$, depending on the nonlinear transformation we obtain from our system. This range of tap values is consistent with the optimal number of FFE taps ($48$) used in the DSP KK receiver implementation approach. This number is significantly higher than the taps used in previous systems \cite{Argyris2018,Argyris2019}, since the chromatic dispersion, introduced by the $100$ km fiber transmission, is much higher in this task.

\begin{figure}[t]
   \centering
        \includegraphics[width=1\linewidth]{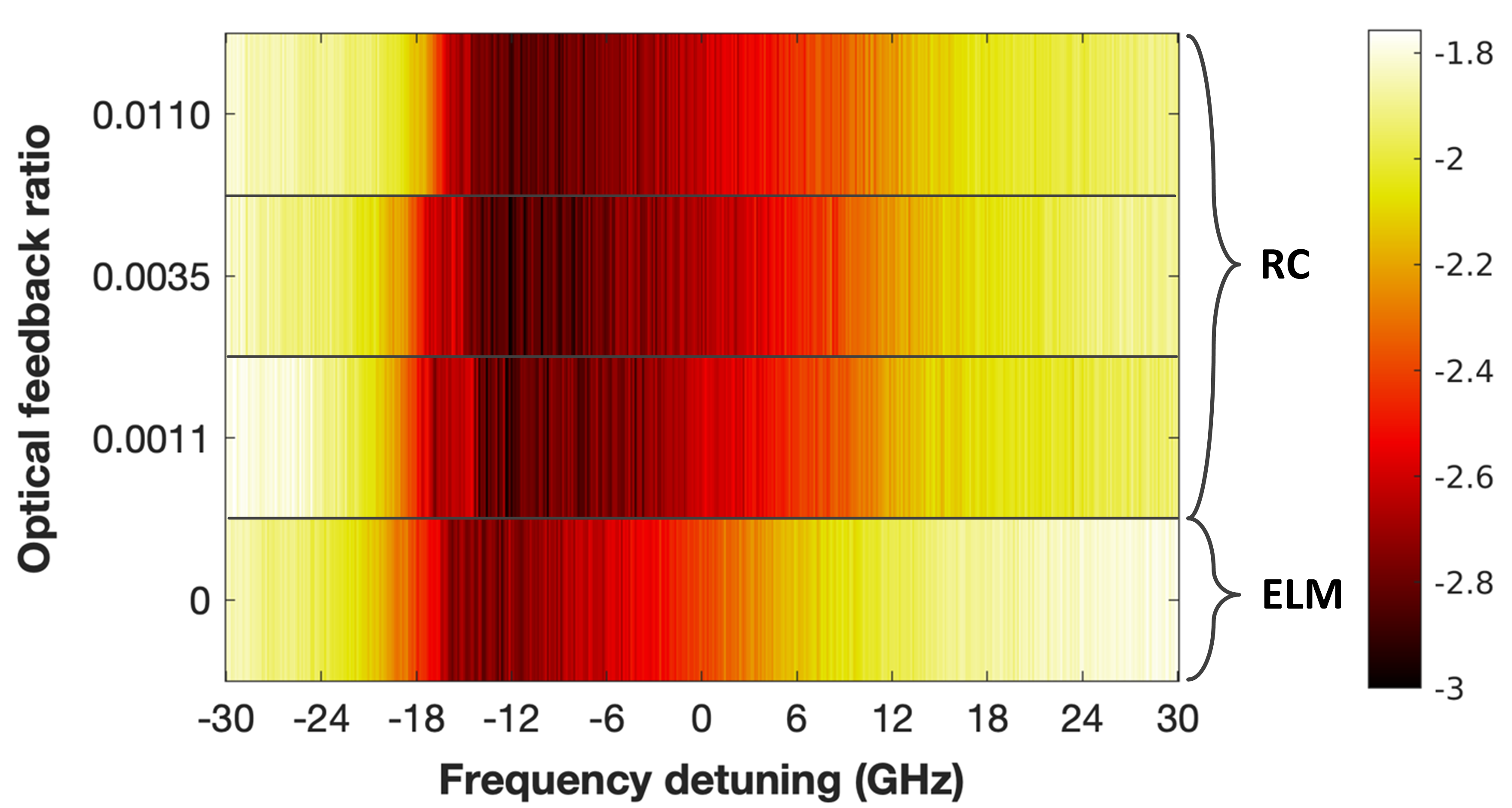}
    \caption{Logarithmic BER performance of data recovery, after TDRC and ELM post-processing, vs. frequency detuning $\Delta f$ and for four different optical feedback ratios of the photonic reservoir. For an open loop implementation that converts RC to ELM, the optical feedback strength is represented by zero.}
    \label{fig:figure5}
\end{figure}

\section{Comparison between TDRC and ELM configurations}
We initially consider a photodetected time-series from the transmission experiment with a high optical signal-to-noise ratio (OSNR $= 35.9$ dB). This high OSNR value was obtained by launching 6dB of average optical power into the fiber transmission line. The photodetected signal is processed with the LR classifier described in the previous section. By optimizing the training data subset size and the number of taps, the lowest error rate we are able to obtain was $\log_{10}(BER)_{REF} = -1.64$. This error rate is well above the hard-decision forward error correction (HD$-$FEC) BER requirement for error-free decoding ($\log_{10}(BER)_{HD-FEC} = -2.42$) \cite{Tzimpragos2014}. At a next step, we use the same time-series to train and evaluate the classification performance of the output response of the TDRC / ELM system in terms of BER, for different frequency detuning $\Delta f$ and feedback conditions of the reservoir.

\subsection{Encoding of one masked symbol per time delay}
Here we investigate the cases (a) and (b) of Fig. \ref{fig:figure4}. The change between a TDRC and an ELM configuration is achieved by opening the feedback loop, via the switch of Fig. \ref{fig:figure2}. We evaluate the BER performance obtained by the two systems, by considering a $125$ MHz resolution mapping versus the frequency detuning $\Delta f$.

The implemented photonic scheme can exhibit a variety of dynamical responses. Under the impact of the dynamical optical injection and for different frequency detuning $\Delta f$ conditions, the system can enter into different regimes of operation \cite{Bueno2017}. These regimes are defined by their optical spectral emission: a fully-locked, a partially-locked, and an unlocked operation. In Fig. \ref{fig:figure5} we show the BER in a logarithmic scale, as calculated for the different operating conditions of the TDRC and the ELM configurations. A first observation is that the lowest error rates are obtained within the same region of frequency detuning ($\Delta f \sim -16$ GHz to $-8$ GHz). This region is associated with the partial locking between the drive and the response laser, consistently with previous works that used this scheme for classification tasks. The lowest logarithmic BER observed for moderate optical feedback ratios (values of $0.0011$ and $0.0035$, in Fig. \ref{fig:figure5}) of the reservoir is $\log_{10}(BER)=-3$. A higher feedback ratio (value of $0.011$, in Fig. \ref{fig:figure5}), that leads the reservoir to operate at the onset of dynamical instabilities, results in a slighter higher error rate. Specifically, the lowest $\log_{10}(BER)$ obtained is $-2.92$. For the ELM consideration, where the feedback loop is open (optical feedback ratio annotated as 0 in Fig. \ref{fig:figure5}), the lowest $\log_{10}(BER)$ is $-2.87$ and is obtained for fewer $\Delta f$ conditions, compared to the TDRC operation. This error rate is only slightly higher than the best TDRC operation. This comparison shows that the symbol connectivity via the optical feedback contributes very limited additional information to the computation.

\begin{figure}[h]
   \centering
        \includegraphics[width=1\linewidth]{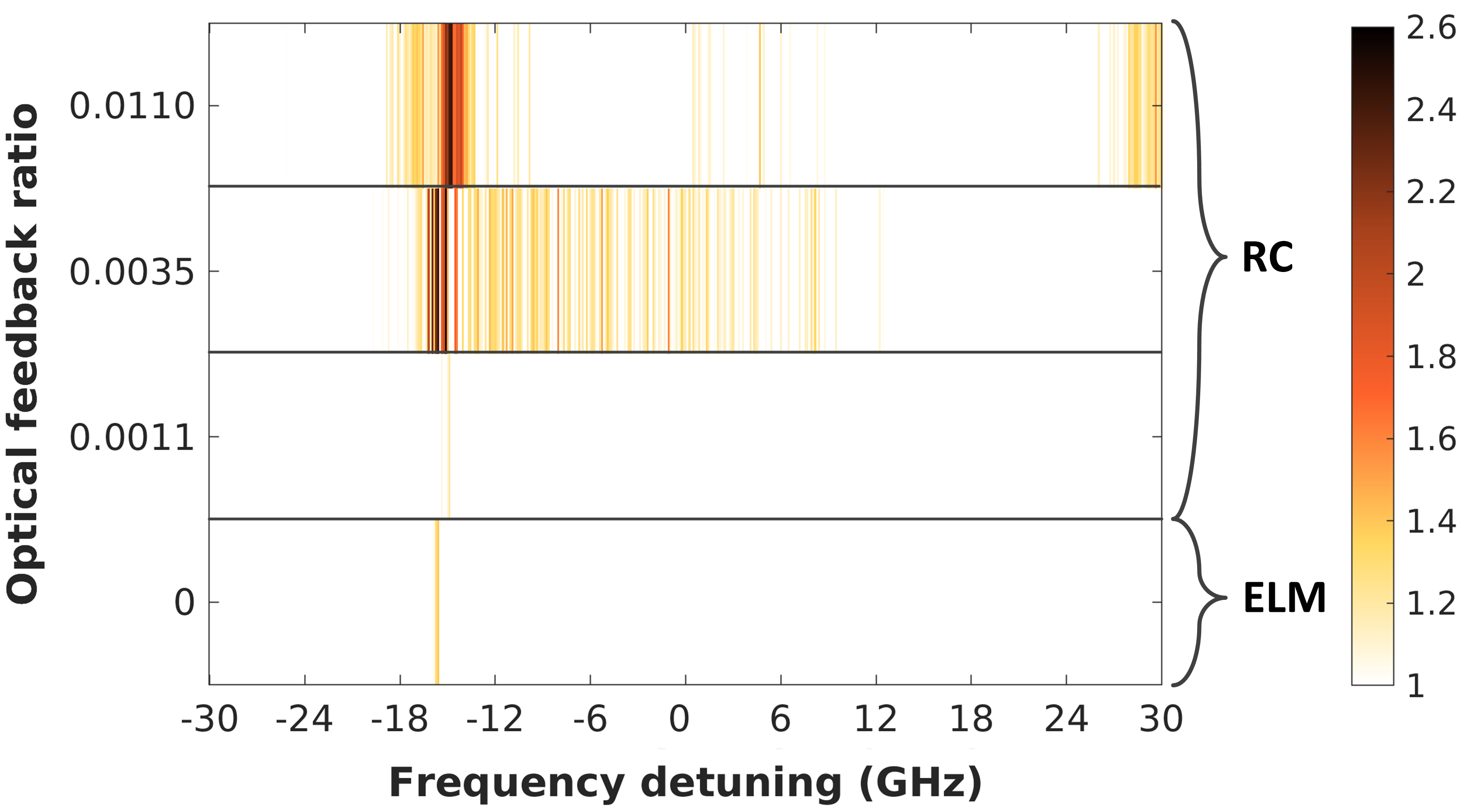}
    \caption{Linear memory capacity ($MC$), as computed experimentally by the photonic TDRC/ELM scheme of Fig. \ref{fig:figure2}, vs. the frequency detuning $\Delta f$ and for four different optical feedback ratios. For an open loop implementation that converts TDRC to ELM, the optical feedback strength is represented by zero.}
    \label{fig:figureMC}
\end{figure}

\begin{figure}[h]
   \centering
        \includegraphics[width=1\linewidth]{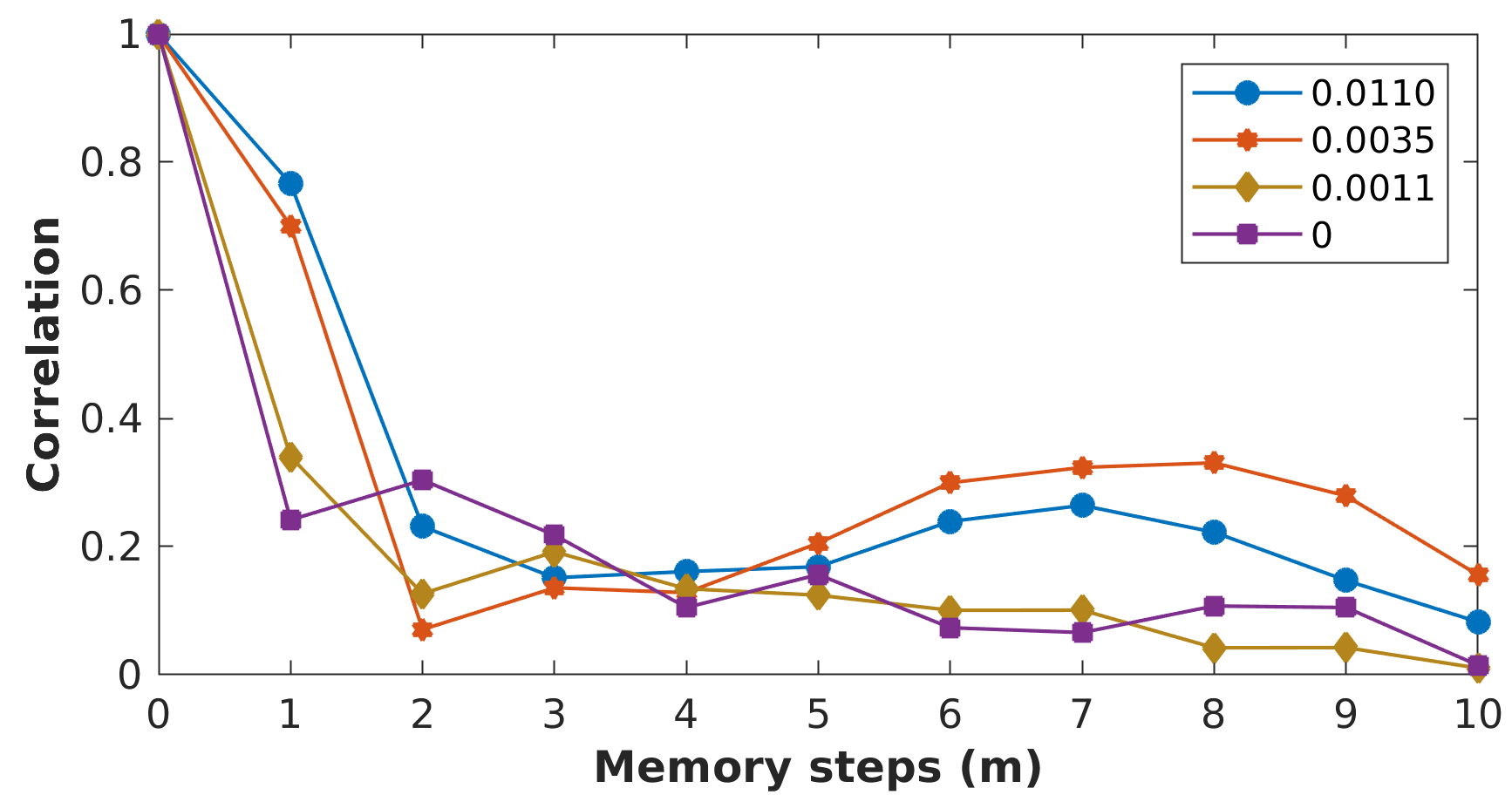}
    \caption{Memory correlation vs. the memory steps, for the highest $MC$ value of each feedback case shown in Fig. \ref{fig:figureMC}.}
    \label{fig:figureCorr}
\end{figure}

To explain this finding, we investigate in more detail the fading memory of the TDRC and compare it with the ELM configuration. In previously investigated data recovery tasks \cite{Argyris2017,Argyris2018,Argyris2019}, the inter-symbol information mixing, due to fiber chromatic dispersion and nonlinearities, was limited from a few up to $20$ neighboring symbols. Here, we investigate a transmission system that by default requires a much higher number of taps ($48$, as demonstrated by the KK receiver implementation and $\sim 40$, when using a linear classifier directly on the transmission output signal). On the other hand, the fading memory of a TDRC depends on the dynamical operation of the reservoir and the number of its virtual nodes. A small number of virtual nodes $-$ as in our investigation $-$ limits the training capability of the system to retain previous information. For the corresponding conditions shown in the task performance of Fig. \ref{fig:figure5}, we calculate the linear short memory capacity ($MC$) of the TDRC \cite{jaeger2002}, using the experimentally obtained responses of the $N = 20$ virtual nodes (Fig. \ref{fig:figureMC}). For the $MC$ calculation, we insert a random sequence of values at the input of our photonic TDRC / ELM system and we train the linear classifier to predict from $m = 1$ and up to $m=10$ memory steps in the past. The operating conditions of the experimental setup are preserved the same as the transmission task. The $MC$ is the sum of the correlations obtained, for all the $m$ steps considered. We see in Fig. \ref{fig:figureMC} that the $MC$ is maximized within the frequency detuning range for which we obtain the lowest decoding error in Fig. \ref{fig:figure5}. The highest value we obtain is $MC = 2.6$ for the two highest feedback ratios. In an ELM configuration, we obtain a significantly lower $MC = 1.6$, as resulted by the absence of the fading memory, associated with the time delay of the reservoir ($\tau$). This difference in the $MC$ values explains the slight improvement in the decoding performance of Fig. \ref{fig:figure5}, when considering TDRC versus the ELM configuration. Still, the above $MC$ values are very low, compared to TDRC systems that consist of a large number of virtual nodes \cite{Bueno2017,lymburn2019,stelzer2020}, but also compared to the tap requirements for the investigated data recovery task. For the highest $MC$ value obtained for each feedback condition of Fig. \ref{fig:figureMC}, we plot the memory correlation versus the different memory steps $m$ (Fig. \ref{fig:figureCorr}). In this graph, we show also $m = 0$, to prove that the classifier is well-trained for the aligned-in-time random sequence. For $m = 1$, higher feedback values of the TDRC favor a higher memory correlation, but for $m > 1$ the memory correlation is less than $0.35$ in all feedback cases. In conclusion, the presence of the feedback loop adds limited internal fading memory to a system with only $N = 20$. For this reason, the TDRC and the ELM configurations show similar performance on the investigated transmission task.

\subsection{Encoding of subsequent masked symbols}
Here we investigate the input encoding cases (c) and (d) of Fig. \ref{fig:figure4}, where the subsequent symbols are now connected additionally through the inertia of the transient states. We follow the same procedure to identify the lowest logarithmic BER, as described in the previous section. The value obtained for the TDRC operation is now $\log_{10}(BER)=-3.16$, while for the ELM operation is $\log_{10}(BER)=-3.12$. We see that the connectivity between neighboring symbols through inertia improves slightly the data recovery performance, compared to the cases of (a) and (b). In the case (c) of the TDRC configuration, the feedback loop introduces a connectivity every $n = \tau/\tau_{m} = 24.5$ ns $/ 1.25$ ns $= 19$ masked symbols. By eliminating this connectivity, in the case (d), we see that this has no impact on the computation performance.

\subsection{ELM for data recovery}
From the previous section we come to the following conclusions: 
\begin{enumerate}
    \item[(a)] When the reservoir's fading memory is significantly shorter than the temporal extent of the inter-symbol interference, the data recovery performance is not enhanced by the TDRC scheme.
    \item[(b)] When operating the nonlinear physical system in a transient regime, the inertia of the transient states provides connectivity between neighboring symbols. This further improves the data recovery performance and is more important than the external cavity connectivity through the optical feedback.
\end{enumerate} 

\begin{figure}[t]
   \centering
        \includegraphics[width=0.98\linewidth]{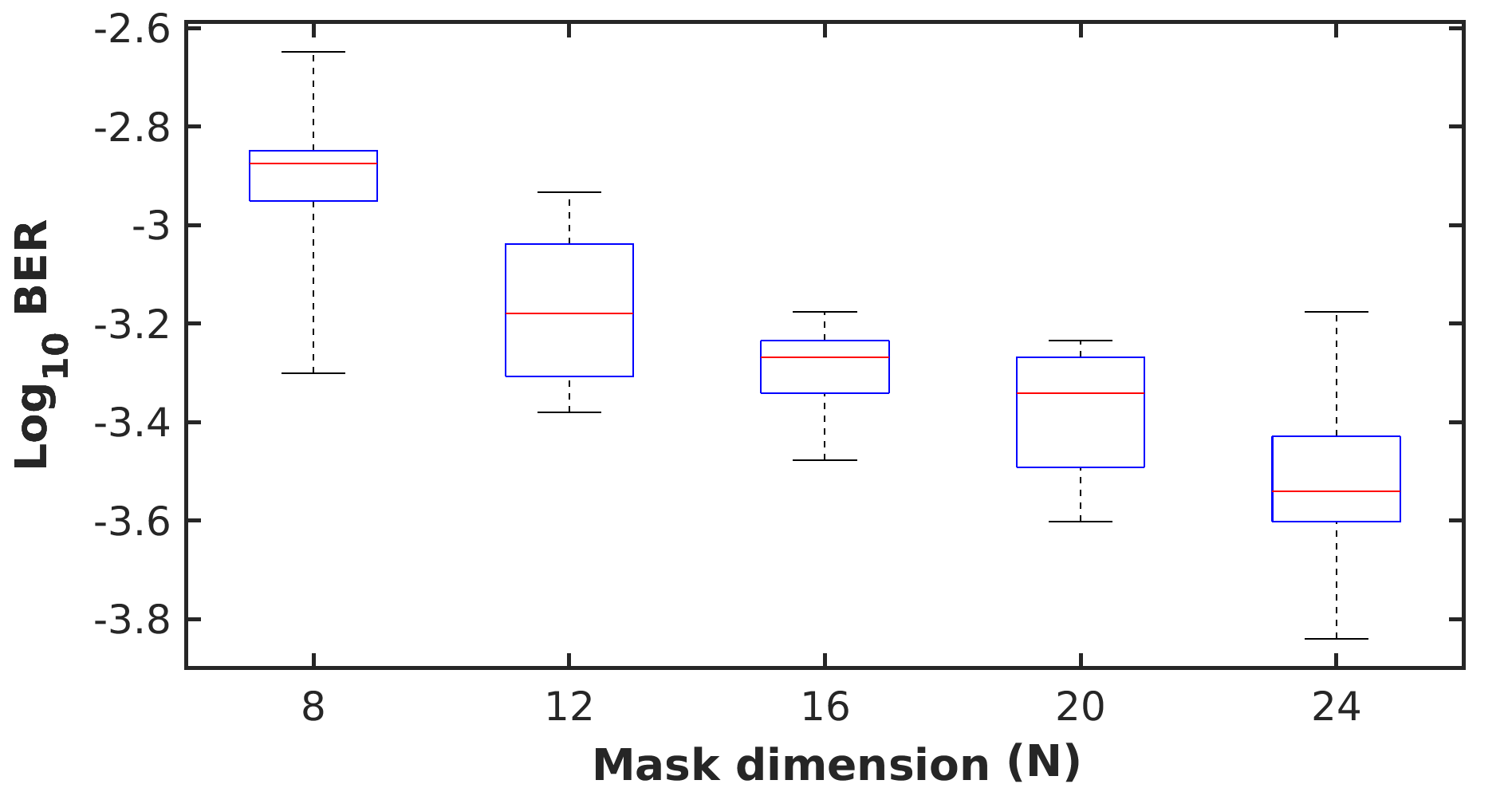}
    \caption{Logarithmic BER performance of data recovery after ELM post-processing vs. the mask dimensionality used for the representation of each symbol, and for 8 different masking sequences. The red line represents the median performance of the various masking sequences, the blue boxes enclose the $75^{th}$ and $25^{th}$ percentiles, while the horizontal black lines represent the maximum and minimum error rates.}
    \label{fig:figure6}
\end{figure}

Thus, we focus on the ELM operation, which in parallel allows us to speed up the computational speed of our system. In absence of the feedback loop, the computational speed depends only on the transient duration ($\theta$) and the dimension of the masking sequence ($N$). Here, we investigate the impact of the mask sequence on the data recovery performance. Fig. \ref{fig:figure6} shows the decoding performance of the ELM topology when considering different dimensionality expansion of the symbols through the masking sequence. For all the studied sizes we consider 8 different masking sequences drawn from a random uniform distribution. We validate what has been also shown in previous investigations \cite{Appeltant2014,Nakayama2016,Kuriki2018,Argyris2021}: the masking sequence plays a decisive role in the task performance. The difference in performance can be half an order or more, in terms of BER. To take full advantage of the computational capability of our ELM topology, a pre-selection of the masking sequence is necessary. In Fig. \ref{fig:figure6} we also observe a significant performance improvement when increasing the number of transient states that enter into the computation. For the investigated signal with OSNR $= 35.9$ dB, we obtain a BER below the HD$-$FEC threshold, even with $N = 8$. When increasing to $N = 24$, we can even reach a $\log_{10}(BER)$ value of $-3.84$. A higher dimensionality of the nonlinear mapping of the input information results in a better performance. However, this comes at the expense of increasing the processing time per symbol.

\begin{figure}[t]
   \centering
        \includegraphics[width=1\linewidth]{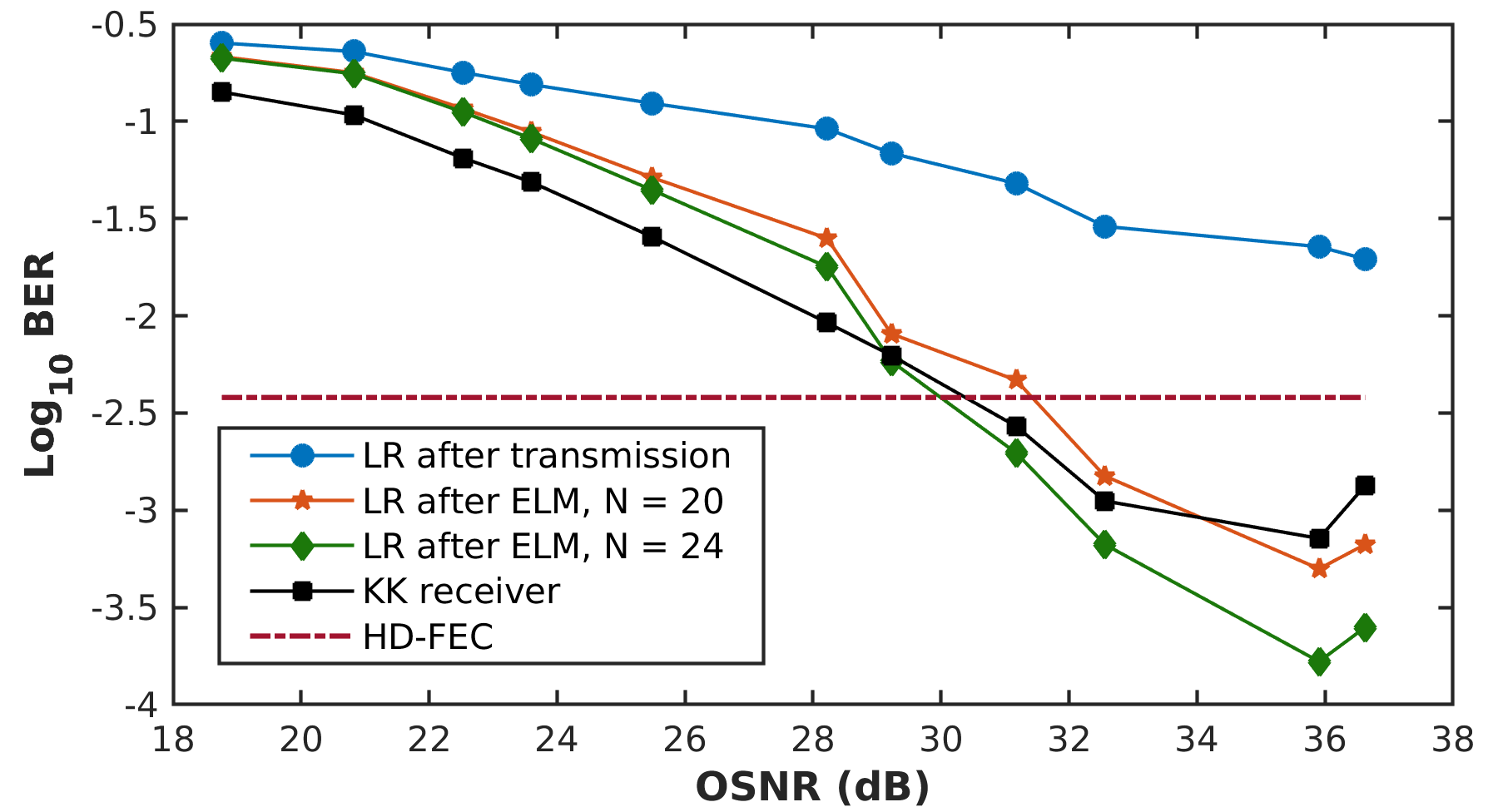}
    \caption{Logarithmic BER performance of data recovery after ELM post-processing, for different values of the OSNR. Blue circles correspond to the LR classifier performance on the data sequence obtained from the fiber transmission system. Orange stars (green diamonds) correspond to the LR classifier performance at the output of the ELM, for the best operating conditions and for mask dimension of \textit{N} = 20 (\textit{N} = 24). Purple rectangles correspond to the result obtained by a DSP-based KK receiver. Dashed red line corresponds to the HD$-$FEC limit for error-free decoding.}
    \label{fig:figure7}
\end{figure}

For the ELM configuration and by using as a masking sequence the one that gives the lowest error rate, we evaluate the data streams from the transmission system of Fig. \ref{fig:figure1} versus their OSNR. In a benchmarking comparison, we evaluate the performance of the same LR classifier used for the ELM topology, by introducing directly the signal obtained from the fiber transmission experiment. In the latter case, only two samples per symbol are used in the classifier. The classifier performance for the different signal OSNRs is shown in Fig. \ref{fig:figure7} (blue dots). The lowest logarithmic BER value is $-1.6$ and is obtained for OSNR $= 36.6$ dB. The performance of the same classifier, when considering the ELM processing stage with a signal dimension expansion of $N = 20$ and $N = 24$ is also shown in Fig. \ref{fig:figure7} (orange stars and green diamonds, respectively). For each calculation, we have optimized the ELM operation to that value of frequency detuning $\Delta f$ that provided the lowest error. When considering an ELM processing with $N = 24$, the lowest $\log_{10}(BER)$ we obtain is $-3.8$, for OSNR $= 35.9$ dB, which is a BER improvement of more than two orders of magnitude compared to the benchmark performance. At this OSNR, the signal suffers not only from linear effects related to chromatic dispersion but also from significant nonlinear distortion, attributed to self-phase and cross-phase modulation. The HD$-$FEC BER limit is achieved for OSNR values above 30 dB. By reducing the ELM dimensionality to $N = 20$, the OSNR for which we achieve the HD$-$FEC BER limit increased to 31.5 dB. However, in this case, the computational speed increases by $20\%$. Finally, in Fig. \ref{fig:figure7} we also compare the ELM performance with the KK receiver, which was implemented via DSP, as presented in \cite{Shi2019}. For OSNR values above 29 dB, we find a clear advantage of the experimental ELM topology with $N=24$. This offers lower error rates, despite all the noise sources that are present in this hardware implementation and affect its overall performance. 

\section{Conclusion}
We demonstrated an experimental ELM system, capable to provide efficient data recovery from a 56 GBaud PAM-$4$ DWDM transmission link over 100 km. In a comparison with a TDRC scheme, we showed that the ELM performs almost equally while simplifying the configuration and eliminating the possible computational speed limitation imposed by the time delay of the TDRC. We have been able to obtain experimentally a BER performance that leads to error-free decoding, for signal OSNRs above 30 dB. For these OSNR conditions, we also showed that this hardware processing implementation performs even better than a DSP-implemented KK receiver. 

\ifCLASSOPTIONcaptionsoff
  \newpage
\fi
\bibliographystyle{IEEEtran}
\bibliography{JLTsubm}

\end{document}